\documentstyle[12pt]{article}
\def\rr{{{\bf r}}}
\def\rrp{{{\bf r^\prime}}}
\def\nup{{{n_\uparrow}}}
\def\ndo{{{n_\downarrow}}}

\begin{document}

\title{ Equilibrium sizes of jellium metal clusters in the stabilized
spin-polarized jellium model}
\author{M. Payami }
\maketitle
\begin{center}
{\it Center for Theoretical Physics and Mathematics,
Atomic Energy Organization of Iran,\\ P.~O.~Box 11365-8486, Tehran, Iran}
\end{center}

\date{\today}
\begin{abstract}
We have used the stabilized spin-polarized jellium model
 to calculate
the
equilibrium sizes of metal clusters.
Our self-consistent
calculations in the local spin-density approximation show that for an
$N$-electron cluster, the equilibrium
is achieved for a configuration in which the difference in the numbers of
up-spin and down-spin electrons is zero or unity, depending on the total
number of electrons.
 That is, a configuration in
which the spins
are maximally compensated.
This maximum spin-compensation results in both the alternation in the average
distance between the nearest neighbor ions and the odd-even alternations in the
ionization energies of alkali metal clusters, in a good agreement with the
molecular dynamics findings and the experiment.
These suggest a realistic
and more accurate method
for calculating the properties of metal clusters in the context of jellium
model than previous jellium model methods.

\end{abstract}
71.15.Hx, 71.15.Mb, 71.15.Nc


\newpage

\section{Introduction}
\label{sec1}

The simplest model used in theoretical study of the properties of simple metal
clusters is jellium model (JM) with spherical geometry.
  \cite{ekardt,knight,brack93} In this model, the ions are replaced by a
uniform positive charge background sphere of density $n=3/4\pi r_s^3$ and
radius $R=(zN)^{1/3}r_s$ where $z$, $N$ and $r_s$ are the valence of the atom,
the number of constituent atoms of the cluster and the bulk value of the
Wigner-Seitz (WS) radius of the metal, respectively. This model can be useful
only when the pseudopotentials of the ions do not significantly affect the
electronic structure. But, it is well-known that the JM has some drawbacks.
\cite{lang70,ashlang67} Keeping the simplicity of the JM, the stabilized
jellium model (SJM) of Perdew {\it et al.}\cite{pertran} has lifted the
essential deficiencies of the JM and significantly improved the results of
calculations. In the SJM, the bulk metal (which is spin-unpolarized in the
absence of external magnetic field) has been made stabilized through the
introduction
of a pseudopotential and fixing its core radius by a value that makes the
pressure on the unpolarized bulk system to vanish.
However, applying the SJM in the framework of rigid jellium
background \cite{paybraj93} may
be suitable for closed-shell clusters of which the spin polarization of the
valence electrons vanishes. It is well-known that, for example, the bond-length
of a diatomic molecule depends on the relative orientations of the valence
electrons. Hence, considering a metal cluster as a large molecule, or the bulk
metal as a huge molecule, one should take account of the volume change due to
spin polarization. We therefore, expect that the spherical jellium radius
should
be different for an $N$-atom cluster with two different spin configurations.
These facts led us to consider a stabilized jellium model in which the spin
degrees of freedom be present.
Stabilizing the jellium system with
non-zero spin polarization, $\zeta$, for the valence electrons resulted in the
stabilized spin-polarized jellium model (SSPJM).\cite{pay98}
Here $\zeta=(n_\uparrow-n_\downarrow)/(n_\uparrow+n_\downarrow)$ and
$n_\uparrow$, $n_\downarrow$ are the spin-up and spin-down electron densities,
respectively. The SSPJM can be applied to metal clusters in two different ways.
The first method, which has been used in Ref.[\ref{pay98}], exploits the fact
that the bulk metal expands as $\zeta$ increases. We call that method as
SSPJM1 throughout this paper. In that method,
the jellium sphere radius is taken as
 $R(\zeta)=(zN)^{1/3}\bar{r}_s(\zeta)$. For $\bar{r}_s(\zeta)$ we had taken
$\bar{r}_s(\zeta)=\bar{r}_s(0)+\Delta r_s(\zeta)$ in which $\bar{r}_s(0)$ was
the observed
value of the bulk WS radius of the metal and $\Delta r_s$ was obtained by the
application of the local spin-density approximation (LSDA) to the infinite
electron gas system.
However, in our phenomenological accounting of the volume change, the core
radius of the pseudo-potential for the electron-ion interaction has been
considered as a parameter, which becomes
polarization dependent as we force the pressure of the polarized bulk system
to vanish.
Using that scheme, we had calculated the energies of different metal
clusters, both neutral and singly ionized, for different spin configurations
and had shown that instead of Hund's first rule for the ground state, the
maximum spin-compensation (MSC) rule was governing.\cite{pay98}
The MSC
property which originates here from the polarization dependence of the core
radius, leads to the odd-even alternations in the ionization
energies that
had been observed experimentally in the alkali metal clusters.\cite{deheer93}
On the other hand, if one assumes a fixed, polarization-independent form for
the electron-ion interaction, the MSC property will be realized only for
non-spherical geometries of the jellium background.

 We have recently shown\cite{pay99} that it is not always necessary for
a finite spherical jellium system
to increase its size as the polarization, $\zeta$, is increased. This can be
explained by considering the fact that for an open-shell cluster if one
increases the spin polarization from
the possible minimum value consistent with the Pauli exclusion principle, one
should make a spin-flip in the last uncomplete shell. Because of high
degeneracy for the spherical geometry, this spin-flip in the
last shell does not change appreciably the kinetic energy contribution to the
total energy but changes appreciably the exchange-correlation energy which in
turn gives
rise to a deeper effective potential that makes the Kohn-Sham
(KS) \cite{kohnsham}
orbitals more localized and therefore a smaller size for the cluster.
On the other hand, although the SSPJM1 results in better ionization energies
than the SJM \cite{paybraj93} in that
it reproduces the odd-even alternation, it always predicts incorrect cluster
sizes. That is, in the SSPJM1, the equilibrium $r_s$ for a cluster is taken to
be greater or equal to the bulk value of $r_s$ ( see Fig. 3 of Ref.
[\ref{pay98}] ), so that it approaches the bulk value from the above; whereas,
the molecular dynamics (MD) results for the average
distance between the nearest neighbor atoms show that the equilibrium $r_s$
value of the neutral clusters are less than the bulk value and it approaches
the bulk value from the below [ see Fig 15(a) of Ref. \ref{roth}].
To incorporate this correct behavior into our SSPJM
calculations, which is the subject of this paper, we proceed parallel to the
work of Perdew {\it et al.}\cite{perdew93} for the spin-polarized case
 and call this method as SSPJM2.
In the SSPJM2, for a given polarization, we first obtain the value of the core
radius of the pseudopotential that stabilizes the bulk system, say $r_c^B$,
and then, using this value of $r_c^B$ in the energy functional of the cluster,
we change the radius of the jellium sphere until the minimum energy is
achieved. Our self-consistent calculations show that the absolute minimum
energy
corresponds to a spin configuration with maximum compensation as in the SSPJM1
case. The equilibrium $r_s$ values corresponding to these minima lie below
the bulk value, reproducing the correct
behavior. These equilibrium $r_s$ values determine the equilibrium sizes of the
clusters.
If we plot the equilibrium $r_s$ value as a function of the number
of constituent atoms in an alkali metal cluster, we see an alternating
behavior, consistent with the MD results.\cite{roth}
 In this paper we have found the equilibrium properties of
neutral and singly ionized Cs, Na and Al clusters of various sizes ($2\le N\le
42$)  using jellium with
sharp boundaries. We have also repeated the SSPJM2 calculations using a jellium
sphere with diffuse boundary. For the sake of comparison, we have derived the
results of the work by Perdew {\it et al.}\cite{perdew93} which is denoted
by SJM1.
Comparing our SSPJM2 results with those of SSPJM1 show that here, the average
energies
per electron and the ionization energies remain more or less the same but
here,
our SSPJM2 calculations show an improvement over the SSPJM1 results for the
equilibrium sizes of the clusters.

In section \ref{sec2} the calculational schemes has been explained. Section
\ref{sec3} is devoted to the results of our calculations and finally, we
conclude this work in section \ref{sec4}.

\section{Calculational Scheme}
\label{sec2}
In the context of the SSPJM, the average energy per valence electron in the
bulk
with density parameter $r_s$ and polarization $\zeta$ is given by\cite{pay98}

\begin{equation}
\varepsilon(r_s,\zeta)=t_s(r_s,\zeta)+\varepsilon_{xc}(r_s,\zeta)+\bar
w_R(r_s,r_c)+\varepsilon_{\rm M}(r_s),
\label{eq1}
\end{equation}
where $t_s$ and $\varepsilon_{xc}$ are noninteracting kinetic energy and
exchange-correlation energies per electron, respectively. $\bar w_R$ is the
average value (over the WS cell) of the repulsive part of the Ashcroft empty
core\cite{ash66} pseudopotential,

\begin{equation}
w(r)=-\frac{2z}{r}+w_R,\;\;\;\;\;w_R=+\frac{2z}{r}\theta(r_c-r),
\label{eq1a}
\end{equation}
and is given by
$\bar
w_R=3r_c^2/r_s^3$. In Eq.(\ref{eq1a}), $z$ is the valence of the atom, and
$\theta(x)$ is the ordinary step function which assumes the value of unity for
positive arguments, and zero for negative values. The core radius, $r_c$, will
be fixed by setting the pressure of the bulk system equal to zero at
equilibrium density $\bar{n}(\zeta)=3/4\pi\bar{r}_s^3(\zeta)$.
In Eq.(\ref{eq1}),
 $\varepsilon_{\rm M}$ is the average Madelung energy,
$\varepsilon_{\rm M}=-9z/5r_0$. Here, $r_0$ is the radius of the WS sphere,
$r_0=z^{1/3}r_s$, and for monovalent metals $z=1$, and for polyvalent metals
we set $z^*=1$ (for details see Ref.[\ref{pertran}]). All equations throughout
this paper
are expressed in Rydberg atomic units. The bulk stability is achieved when
$r_c$ takes a value that makes the pressure of the system with a given
$\zeta$ to vanish at $r_s=\bar r_s(\zeta)$:

\begin{equation}
\left.\frac{\partial}{\partial r_s}\varepsilon(r_s,\zeta,r_c)\right|_{r_s=\bar
r_s(\zeta)}=0.
\label{eq2}
\end{equation}

The derivative is taken at fixed $\zeta$ and $r_c$. Solution of the above
equation gives the bulk value of $r_c$ as a function of $\bar{r}_s$ and
$\zeta$.
Here, $\bar r_s(\zeta)$ is
the equilibrium density parameter for the bulk system with given $\zeta$ and is
evaluated by
\begin{equation}
\bar{r}_s(\zeta)=\bar{r}_s(0)+\Delta r_s^{\rm EG}(\zeta).
\label{eq2a}
\end{equation}
Here, $\bar{r}_s(0)$ takes the observed value for a metal and for Cs, Na, and
Al it takes the values of 5.63, 3.99, and 2.07, respectively. In the second
term of the right hand side, the superscript
``EG'' refers to the electron gas, and $\Delta r_s^{\rm EG}(\zeta)$ is
evaluated by setting the pressure of the electron gas system equal to zero
(see Eq.(19) of Ref.[\ref{pay98}]).
The solution of Eq.(\ref{eq2}) at equilibrium density gives the bulk value of
the
core radius:

\begin{equation}
r_c^B(\bar r_s,\zeta)=\frac{\bar r_s^{3/2}}{3}\left\{
-2t_s(\bar r_s,\zeta)-\varepsilon_x(\bar r_s,\zeta)+\bar r_s
\left(\frac{\partial}{\partial
\bar r_s}\right)_\zeta\varepsilon_c(\bar r_s,\zeta)-\varepsilon_M(
\bar r_s)\right\}^{1/2}.
\label{eq2b}
\end{equation}
Now, using $r_c^B$ in the SSPJM energy functional of a cluster [ Eq.(20)
of Ref.[\ref{pay98}]], we obtain the SSPJM2 energy as

\begin{eqnarray}
E_{\rm SSPJM2}[\nup,\ndo,n_+]&=&
E_{\rm JM}[\nup,\ndo,n_+]+(\varepsilon_M(r_s)+\bar w_R(r_c^B,r_s
))\int d\rr\;n_+(\rr) \nonumber \\
  &&+\langle\delta v\rangle_{\rm WS}(r_c^B,r_s)\int
d\rr\;\Theta(\rr)[n(\rr)-n_+( \rr)],
\label{eq2c}
\end{eqnarray}
where
\begin{eqnarray}
E_{\rm JM}[\nup,\ndo,n_+]&=&T_s[\nup,\ndo]+E_{xc}[\nup,\ndo] \nonumber\\
&&+\frac{1}{2}\int d\rr\;\phi([n,n_+];\rr)[n(\rr)-n_+(\rr)]
\label{eq2d}
\end{eqnarray}
and
\begin{equation}
\phi([n,n_+];\rr)=2\int d\rrp\;\frac{[n(\rrp)-n_+(\rrp)]}{\mid\rr-\rrp\mid}.
\label{eq2e}
\end{equation}
In Eq. (\ref{eq2c}), $\langle\delta v\rangle_{\rm WS}$ is the average of the
difference potential over the Wigner-Seitz cell and the difference potential,
$\delta v$, is defined as the difference between the pseudopotential of a
lattice of ions and the electrostatic potential of the jellium background.

The first and second terms in the right hand side of Eq.(\ref{eq2d}) are the
non-interacting kinetic energy and the exchange-correlation energy, and the
last term is the Coulomb interaction energy of the system.
In our spherical JM, we have
\begin{equation}
n_+(\rr)=\frac{3}{4\pi r_s^3}\theta(R-r)
\label{eq2f}
\end{equation}
in which $R=(zN)^{1/3}r_s$ is the radius of the jellium sphere, and $n(\rr)$
denotes the electron density at point $\rr$ in space.

Applying the SSPJM2 to an $N$ electron cluster with $N_\uparrow$ up-spin and
$N_\downarrow$
down-spin electrons ($N=N_\uparrow+N\downarrow$) and polarization
$\zeta=(N_\uparrow-N_\downarrow)/(N_\uparrow+N_\downarrow)$, the total energy
becomes
a function of $N$, $\zeta$, $r_s$, and $r_c^B$ where $r_s$ is the density
parameter of the jellium background and $r_c^B$ is given by Eq.(\ref{eq2b}).
The
equilibrium density parameter, $\bar r_s(N,\zeta)$, for a cluster is the
solution of

\begin{equation}
\left.\frac{\partial}{\partial r_s}E(N,\zeta,r_s,r_c^B)\right|_{r_s=\bar
r_s(N,\zeta)}=0.
\label{eq3}
\end{equation}

Here again, the derivative is taken at fixed values of $N$, $\zeta$, and
$r_c^B$.
For an $N$-electron
cluster, we have solved the KS equations\cite{kohnsham}
self-consistently for various spin configurations and $r_s$ values and obtained
the absolute minimum-energy spin configuration and its corresponding density
parameter.


\section{Results}
\label{sec3}

We have applied the SSPJM2 to calculate the equilibrium energies and sizes of
different metal clusters. In our calculations for an $N$ electron cluster, we
have solved the KS equations for all possible spin configurations $0\le\zeta\le
1$ and obtained the minimum values of energies and corresponding $r_s$ values
of each configuration. The self-consistent
calculations for Cs, Na, and Al with $2\le N\le 42$ show that the absolute
minimum-energy configuration obeys the MSC rule and the equilibrium $r_s$ value
for the cluster, $\bar r_s(N,\zeta)$, is less than the bulk value because, for
small clusters the ratio of surface to volume energies become comparable and
the surface tension compresses the cluster. This effect is known as
self-compression.\cite{perdew93}
In Fig. \ref{fig1} we have compared the
equilibrium $r_s$ values of ``generic clusters'', JM1 (see Ref. \ref{pay99}),
with the SJM1 results which reproduce correct trends.\cite{perdew93}
To clarify the concept of the ``generic cluster'', suppose that
one solves the self-consistent KS-LSDA
equations for a spherical simple JM cluster with jellium radius equal to
$R=N^{1/3}r_s$ and total number of electrons $N$. For a given $N$, these
calculations are performed for different $r_s$ values as well as different
spin configurations until the
equilibrium $r_s$ value, $\bar{r}_s(N,\zeta)$, corresponding to the absolute
minimum-energy configuration is obtained ( See Fig. 4 of Ref.\ref{pay99} ).
Since in the calculations one does not use
any specific parameter corresponding to a certain metal, the result does not
simulate any real cluster, and we call it an $N$-electron ``generic cluster''.
In the limit
of $N\rightarrow\infty$, the infinite generic cluster tends to the electron
gas system for which $\zeta\rightarrow 0$ and $\bar{r}_s\rightarrow 4.18$. As
is seen from
the figure, the equilibrium $r_s$ value for the generic cluster approaches the
bulk value, 4.18, from the above which does not simulate the correct behavior
for a real metal cluster.
It is seen that for the generic clusters the equilibrium values are
greater than the value $r_s=4.18$ whereas the SJM1 predicts values that are
smaller than the bulk value for Na ($r_s=3.99$) in agreement with the
MD findings. This comparison clearly shows that simple JM
gives wrong molecular bond lengths.
We have performed our calculations based on Eq.~(\ref{eq3})
both for jellium with sharp boundary, SSPJM2, and diffuse jellium,
dif-SSPJM2. For our diffuse jellium calculations we have used the background
density with the radial dependence as\cite{rubio}

\begin{equation}
n_+(r)=\left\{\begin{array}{l}
              n\{1-(R+t)e^{-R/t}[\sinh(r/t)]/r\},\;\;\;r\le R\\
              n\{1-((R+t)/2R)(1-e^{-2R/t})\}R e^{(R-r)/t}/r,\;\;\;r>R,
\end{array}
        \right.
\label{eq4}
\end{equation}
 where $n=3/4\pi r_s^3$ , $R=N^{1/3}r_s$, and $t$ is a parameter related to the
surface thickness. We have chosen
$t=1$ in all our diffuse jellium calculations and then have varied the value of
$r_s$ until the minimum energy is achieved.  Figure \ref{fig2}(a) compares the
equilibrium $r_s$ values of neutral cesium clusters for different sizes. It is
seen that in most cases (rather large clusters for which $\zeta<<1$ ), the
SSPJM2 and
the SJM1 predict the same values for the equilibrium $r_s$; whereas for $N=$
3, 5, and 7 the SSPJM2 predicts larger values. These larger values give rise
to an alternation in the plot. The values obtained from the dif-SSPJM2 lie
below the values obtained from the SSPJM2 and the SJM1.
 Figure \ref{fig2}(b) shows the results
obtained for singly ionized cesium clusters. In this case, we see that the
alternations persist up to $N$=15 and have relatively large amplitudes. For
$N$=3 i.e., singly ionized 4-atom cluster, the value obtained from the SSPJM2
has become larger than the bulk value which is related to the rough
evaluation of $\Delta r_s$. In Fig. \ref{fig3}(a) we have shown the same
quantities for neutral Na clusters. The behavior of SSPJM2 results is the same
as in Fig. \ref{fig2}(a) but, in the case of the dif-SSPJM2 the value for
$N$=5 has become nearly equal to that of $N$=6 and also, the value for $N$=3 is
less
than that of $N$=4 which completely differs from the SSPJM2 results. Also, we
could not find any finite value for $N$=2 case in the dif-SSPJM2. That is, as
much as we decrease the input $r_s$ value, the total energy correspondingly
decreases. This means that the surface tension dominates the internal pressure
and collapses the cluster. Of course, this is not the case in reality and it is
the consequence of the fact that here the surface thickness, $t$, has become
comparable to the cluster radius, $R$.
Figure \ref{fig3}(b) compares the plots of average distance between the nearest
neighbors obtained from our SSPJM2 calculations and the MD calculations of
R\"othlisberger and Andreoni.\cite{roth} In order to estimate the average
distance between the nearest neighbor ions in the cluster, we have assumed a
{\it bcc} structure as in the bulk of Na. Then the shortest distance between
the ions, $d$, is related to the lattice constant, $a$, through
$d=a\sqrt{3}/2$. But, in the {\it bcc} structure for Na, there are two
electrons in a cell and therefore, $a=2\sqrt{\pi/3}\;r_s$.
Combining these two relations results in $d=\sqrt{\pi}\;r_s$. The value
$r_s=3.99$ is appropriate for room temperature ($T=300\;K$) which results in a
value of $d=7.07$ bohrs whereas, for $T=0\;K$ the appropriate value for $r_s$
is 3.93 which gives rise to the value $d=6.96$ bohrs. Therefore, our results
should lie above the MD results [ see Fig 15(a) of Ref. \ref{roth}] because,
the MD calculations have been performed for zero temperature.
In Fig. \ref{fig3}(c) we have shown the plots of equilibrium $r_s$ values for
singly ionized sodium clusters as functions of number of electrons, $N$. The
behavior is more or less the same as neutral one. Figure \ref{fig4}(a) compares
the SSPJM2 results for neutral Al clusters with the results obtained using the
SJM1. Here, we have taken the effective value of $z^*=1$. The diamonds and
squares in the plot show the physical points. The main difference between our
results and the SJM1 is in the size of the jellium atom of Al. In Fig.
\ref{fig4}(b) we have compared the results for singly ionized Al clusters. The
results show some differences for values of $N$ away from shell closings.
Looking
at the above-mentioned figures, we note that in all the three cases of Cs,
Na, and Al the SSPJM2 results in a larger or equal values
for the average distance than the SJM1, and in addition show alternations for
small clusters.

Finally, in Fig. \ref{fig5}(a) we have compared the plots of the total energies
per electron in the two schemes of the SSPJM2 and the SSPJM1 for Na clusters.
It is seen that in the SSPJM2 the energies are relatively lower than those
of the SSPJM1 for smaller clusters but the same for larger ones. We have also
calculated the ionization energies of the clusters using the dif-SSPJM2 and
and compared with the dif-SSPJM1 and experimental values in
Fig. \ref{fig5}(b).
Here also the
odd-even alternations show up themselves in the SSPJM2 as well as in
the dif-SSPJM2 results and the values
obtained are more or less the same as in the dif-SSPJM1 [see Fig. 7(c) of Ref.
\ref{pay98}].
Therefore, the SSPJM2 calculations for simple metal clusters has improved the
previous work, SSPJM1, in that it not only reproduces the odd-even alternations
in the ionization energies, but also it gives correct behavior for the
equilibrium sizes of the clusters.

\section{Summary and Conclusion}
\label{sec4}
In this work we have performed the SSPJM calculations as in the case of
{\it ab initio} molecular structure calculations. That is, we have firstly
calculated the stabilizing core radius of the pseudopotential, $r_c^B$, for
the bulk system with nonzero spin polarization. Then, using this value in the
energy functional of a cluster with given values of $N$ and $\zeta$, the energy
becomes a function of the single variable $r_s$, the density parameter of the
uniform jellium background. Minimizing this function with respect to $r_s$
gives us the equilibrium
$r_s$ value and energy of the cluster with that specified $N$ and $\zeta$.
Our self-consistent KS-LSDA calculations show that the
equilibrium configuration is one in which the spins are maximally compensated
as in our previous findings.\cite{pay98} This maximum spin compensation gives
rise to the odd-even alternations
seen in the experimental ionization energy plot of alkali metal clusters.
Calculating the average
distance between the nearest neighbors of Na clusters, we find a good agreement
between our SSPJM2
results and those obtained from MD calculations. We have therefore improved our
previous SSPJM1 results in that the odd-even property is kept the same as
before
but here, the sizes of the smaller clusters have been predicted correctly.

{\Large\bf Acknowledgements}

The author would like to thank John P. Perdew for reading the manuscript
and the useful discussions on the subject. He also
acknowledges Bahram Payami for providing computer facilities.

\newpage

\begin{figure}
\caption{The equilibrium $r_s$ values in atomic units as functions of the
cluster size $N$. The solid squares correspond to the ``generic clusters''
(JM1)
defined in the text and the large squares correspond to Na clusters using the
method
of Ref.\protect\ref{perdew93}. The dashed and solid lines correspond to the
equilibrium $r_s$ value of the bulk (4.18) in simple JM and to the bulk value
of sodium (3.99), respectively.}
\label{fig1}
\end{figure}

\begin{figure}
\caption{(a) The equilibrium $r_s$ values in atomic units as functions of the
cluster size for cesium clusters obtained from using the three schems
of SJM1, SSPJM2, and dif-SSPJM2. The dashed line corresponds to the bulk
value of $r_s$ for cesium (5.63). (b) Same as (a) but for singly ionized
cesium clusters. }
\label{fig2}
\end{figure}

\begin{figure}
\caption{(a) Same as Fig.\protect\ref{fig2} for neutral Na clusters. The bulk
value is 3.99. (b) The
average distance between nearest neighbors in atomic units for Na clusters. The
squares
correspond to our findings through SSPJM2, appropriate for room temperature
and
the diamonds correspond to the molecular dynamics results at zero temperature.
The dotted line correspond to the bulk value 7.07. (c) Same as (a) for singly
ionized Na clusters.}
\label{fig3}
\end{figure}

\begin{figure}
\caption{(a) The equilibrium $r_s$ values in atomic units for neutral Al
clusters ($z^*=1$, $r_s=2.07$). The diamonds and squares show the physical
points in SSPJM2 and SJM1 schemes, respectively. (b) Same as (a) for singly
ionized Al clusters.}
\label{fig4}
\end{figure}

\begin{figure}
\caption{(a) The total energies per atom of Na clusters in electron volts for
the SSPJM2 and the SSPJM1.
The SSPJM2 results are somewhat lower than those of the SSPJM1 for smaller
clusters. (b) The ionization energies in electron volts for Na clusters in the
dif-SSPJM2 and the dif-SSPJM1 are compared with the experimental
values.\protect\cite{deheer93} }
\label{fig5}
\end{figure}
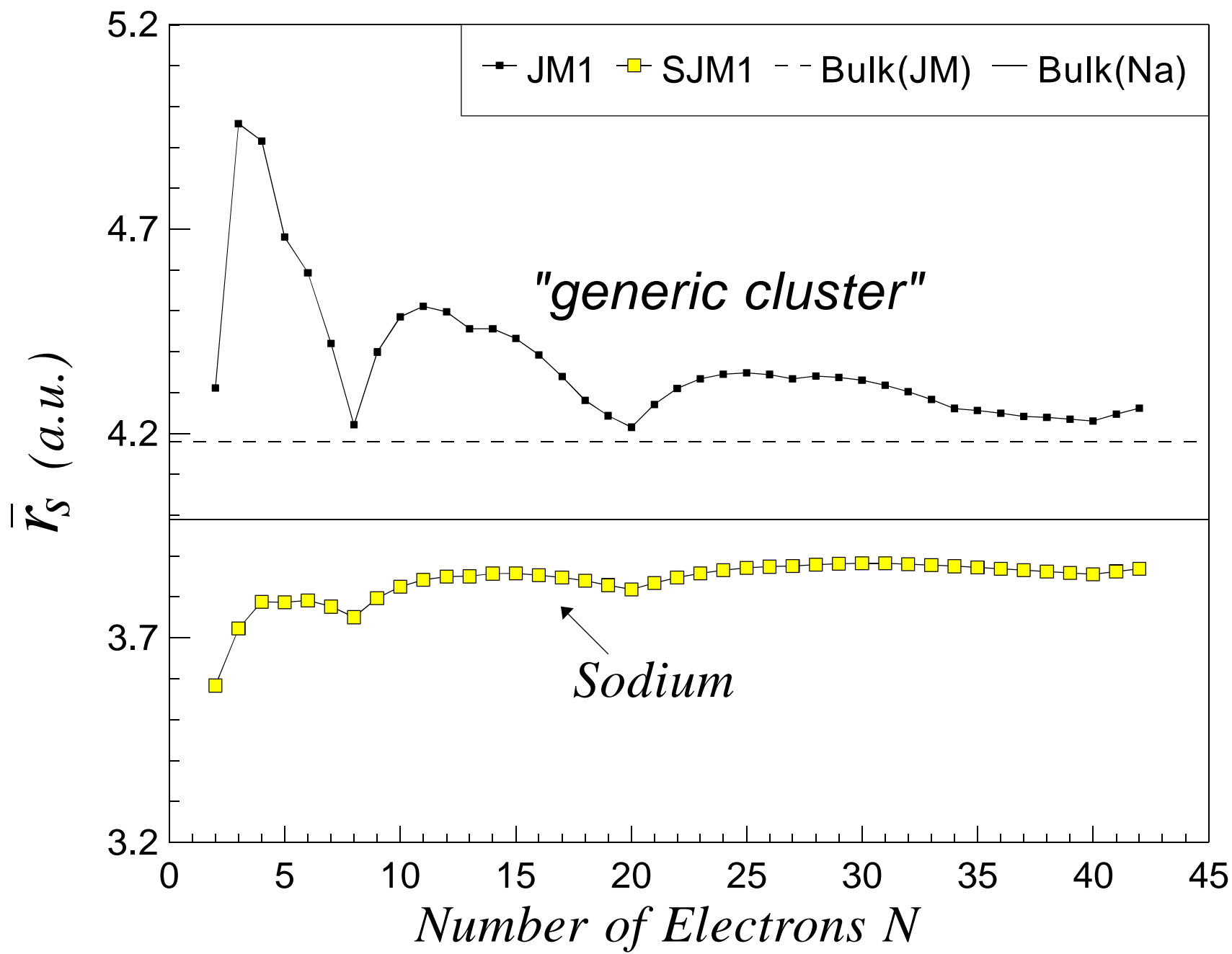
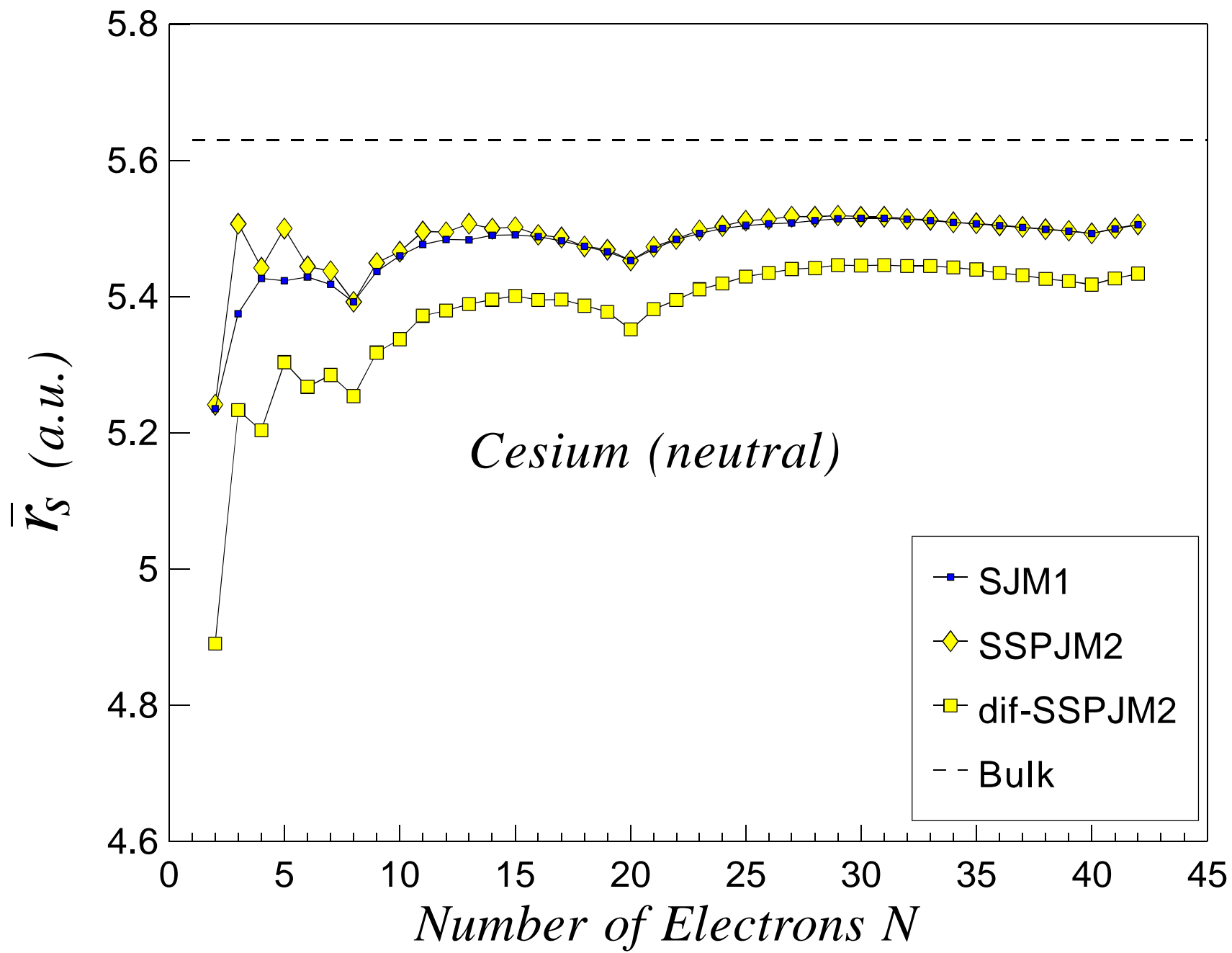
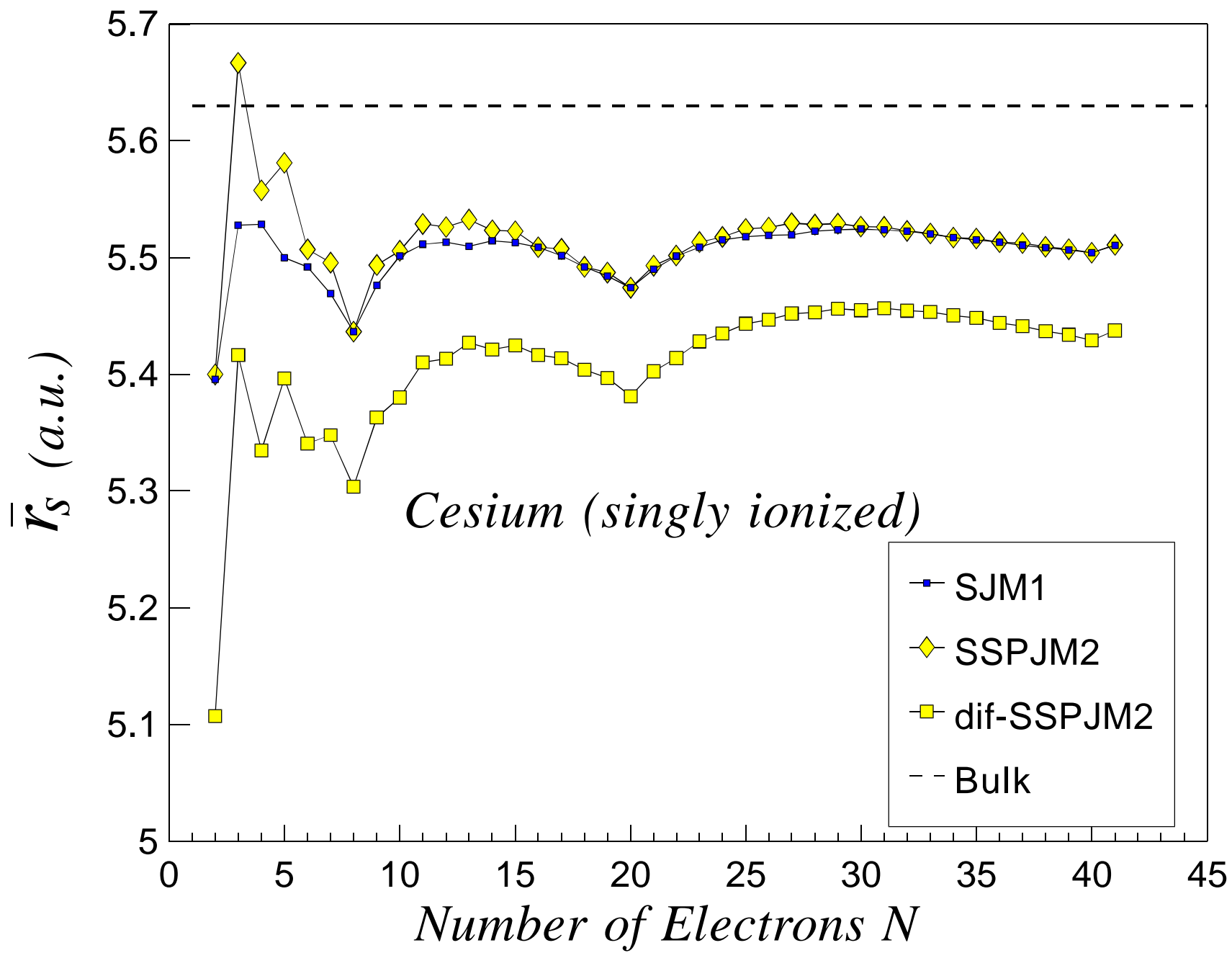
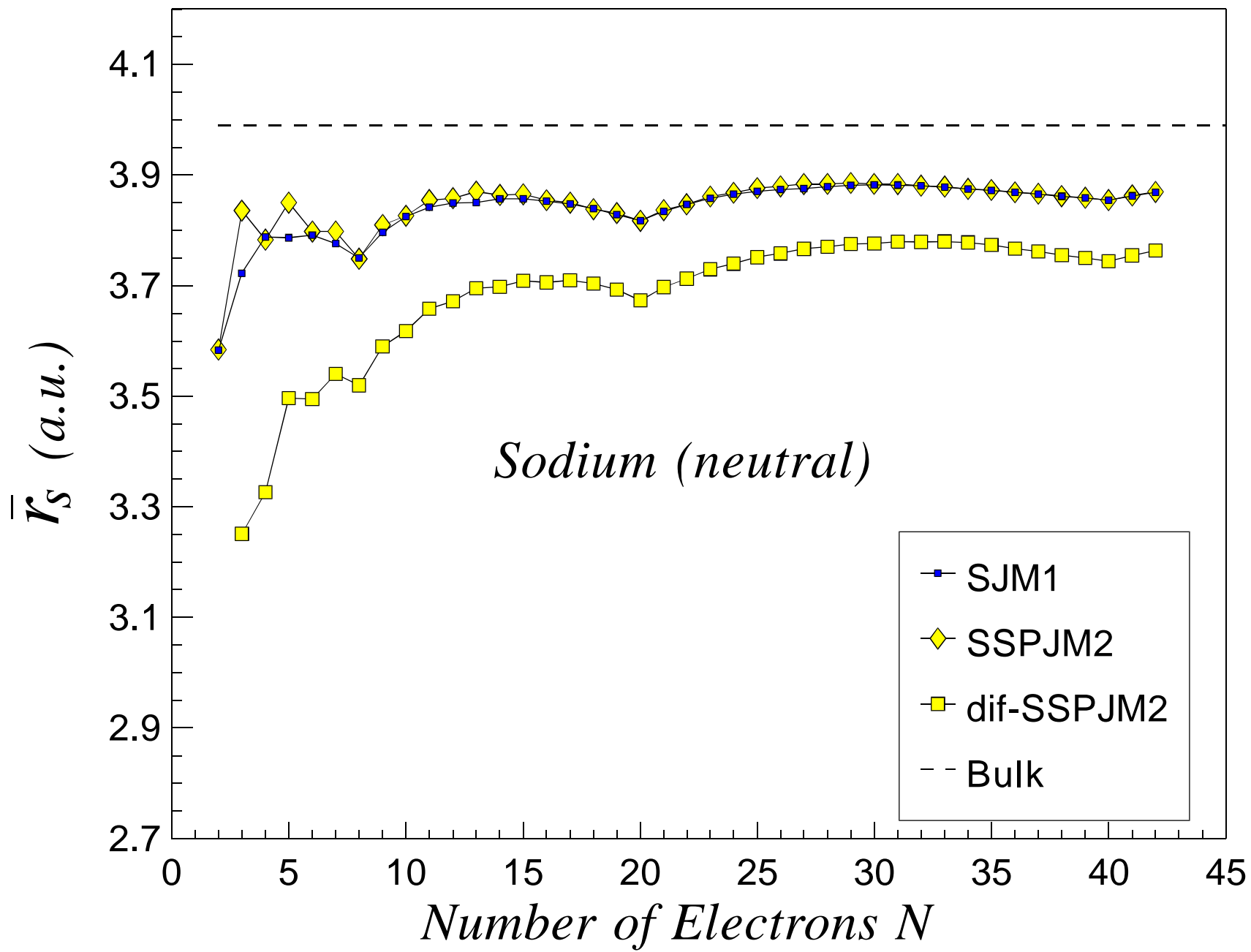
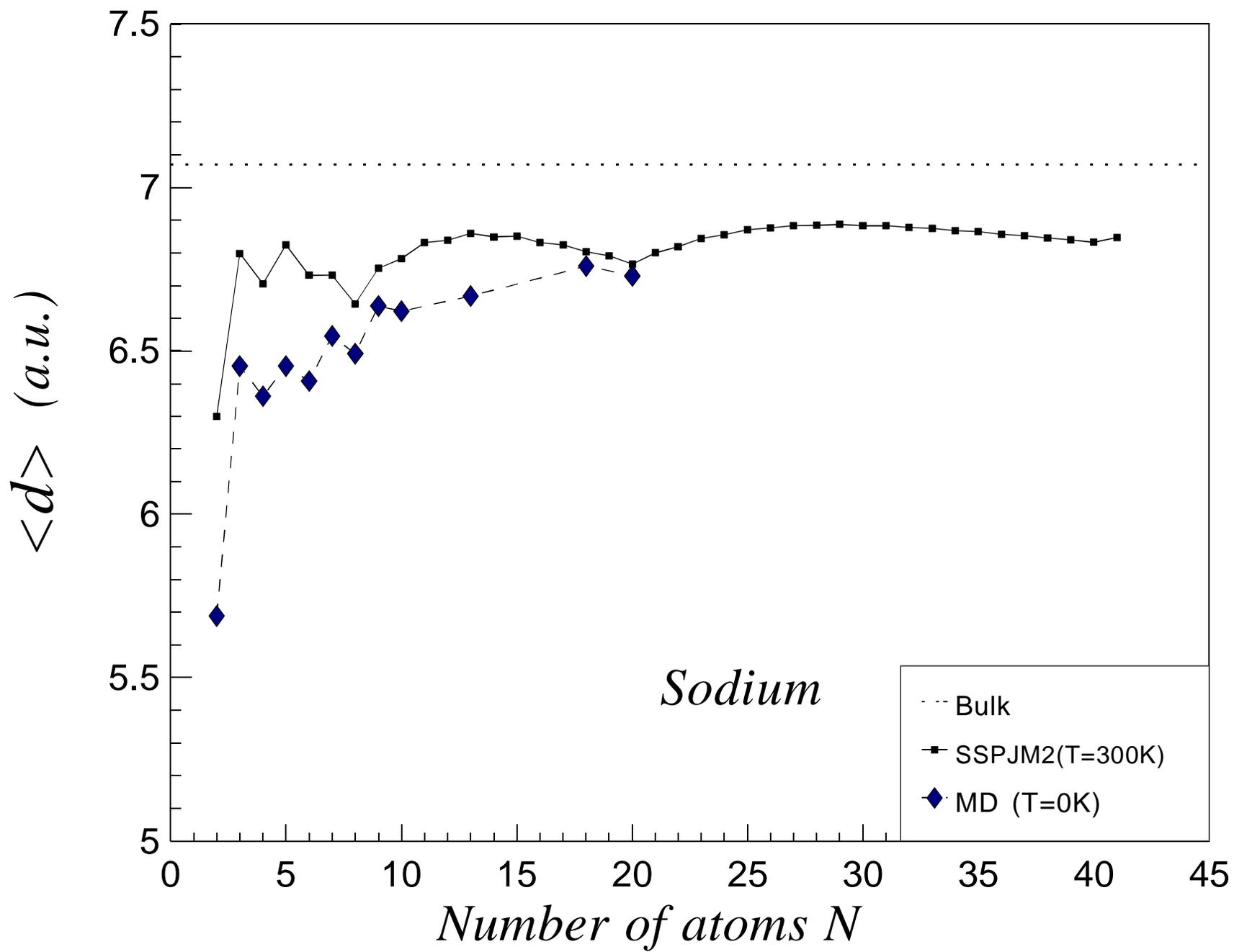
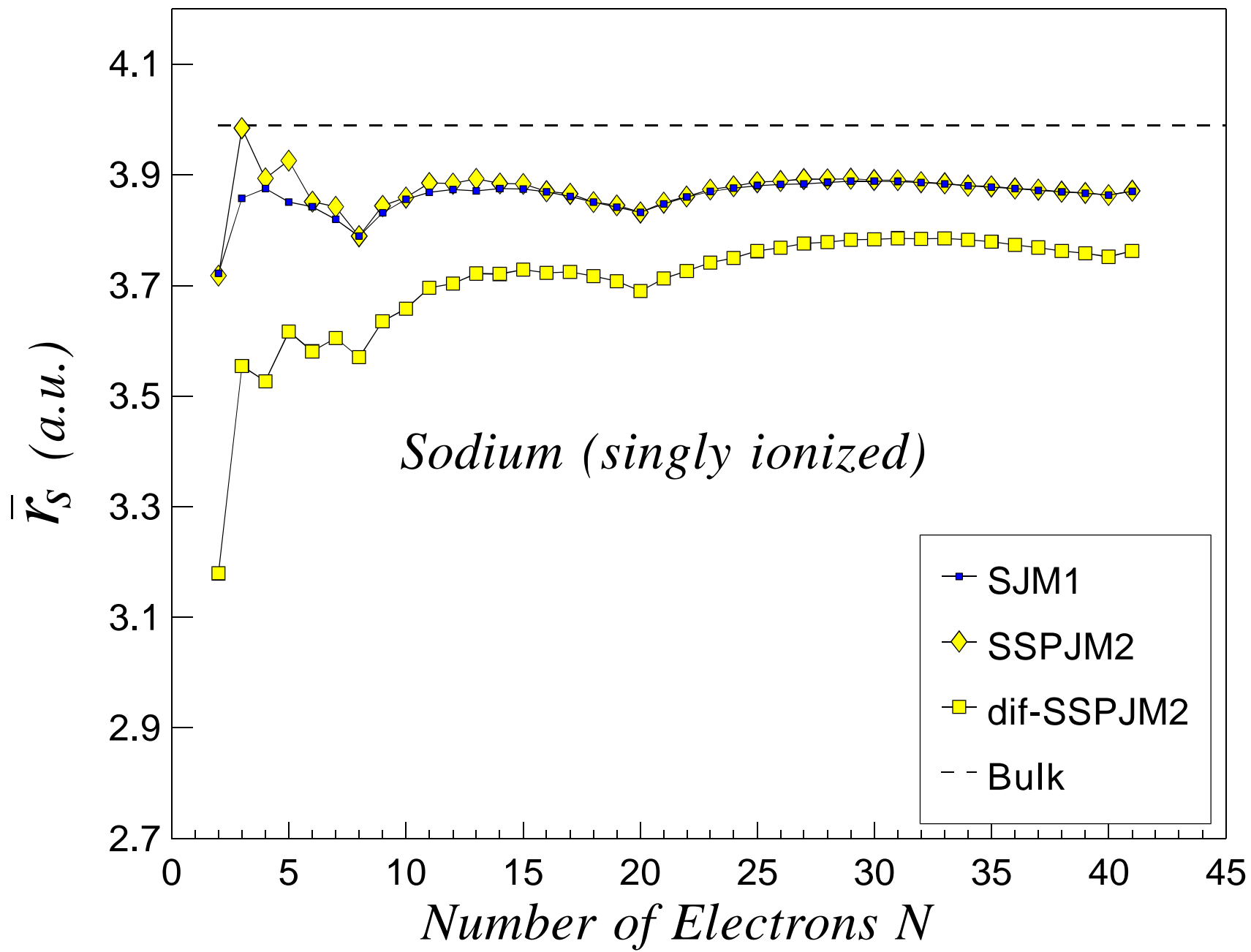
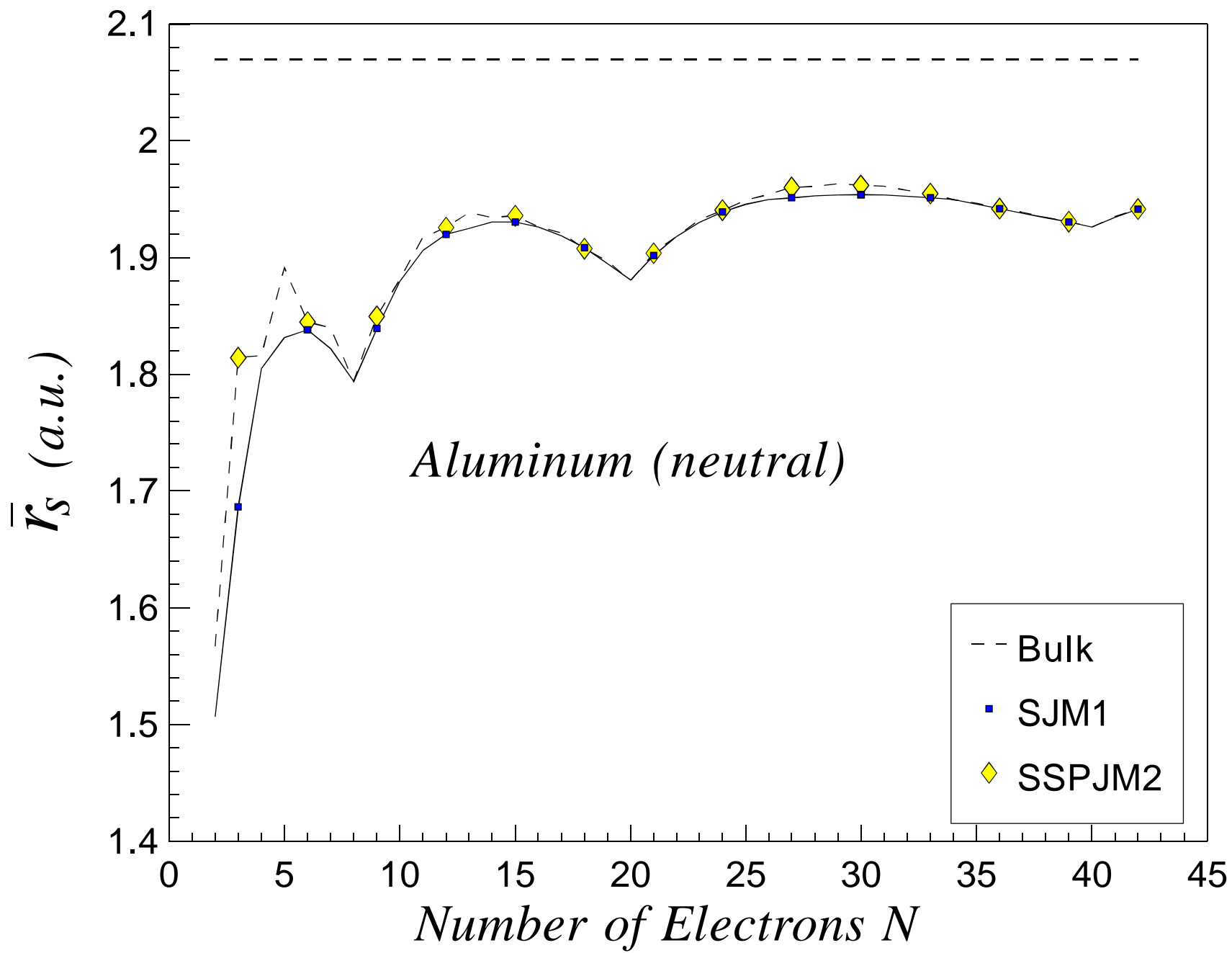
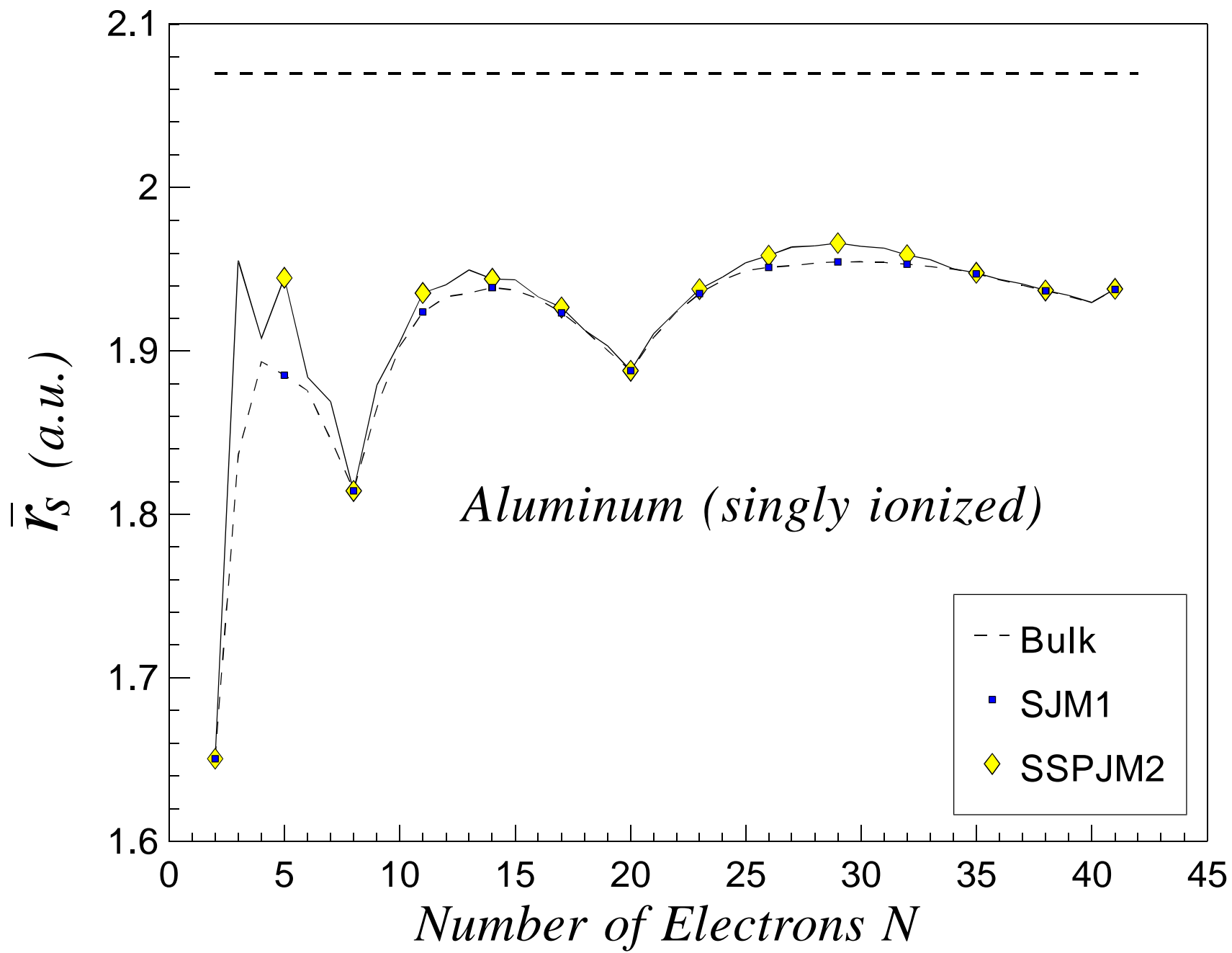
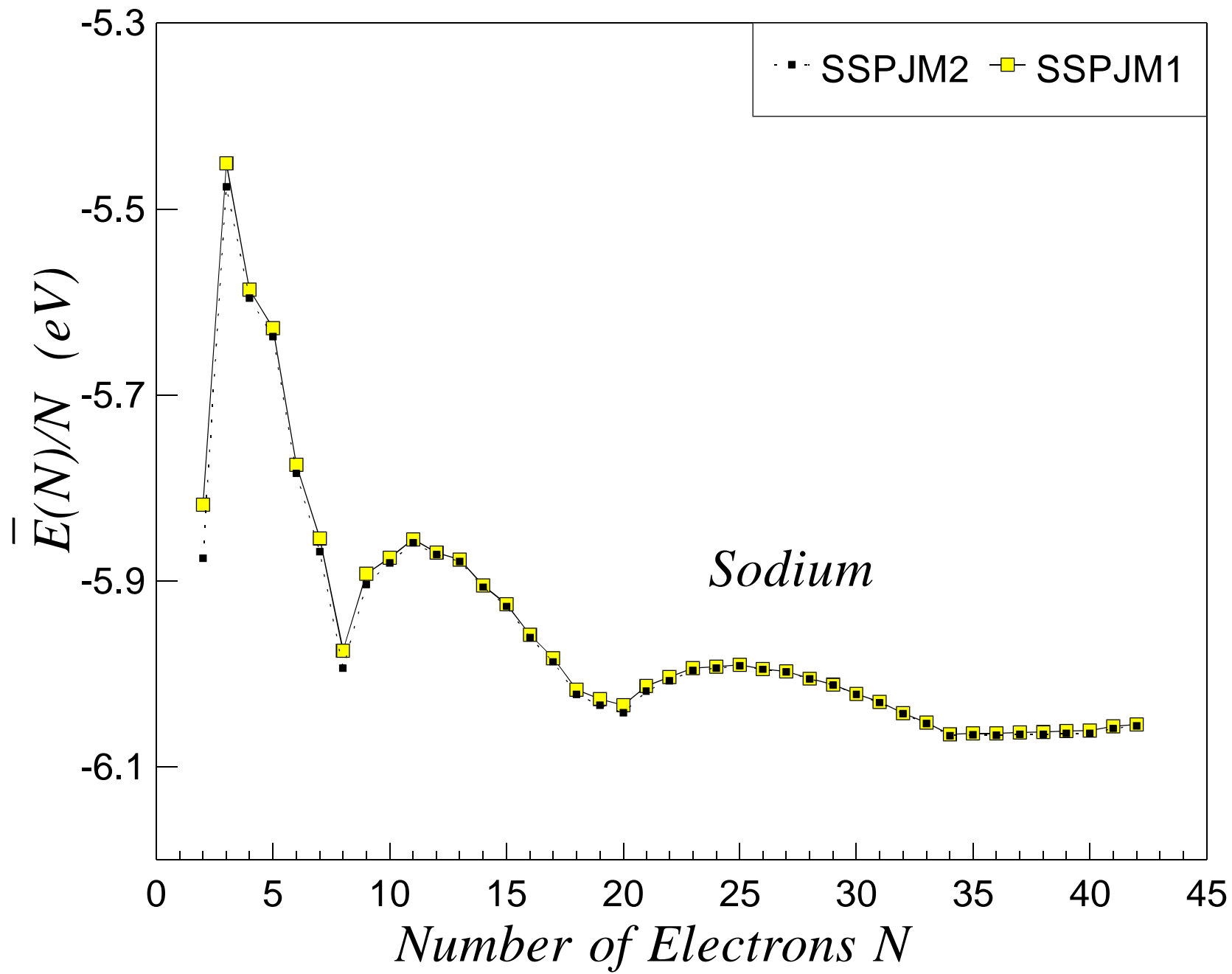
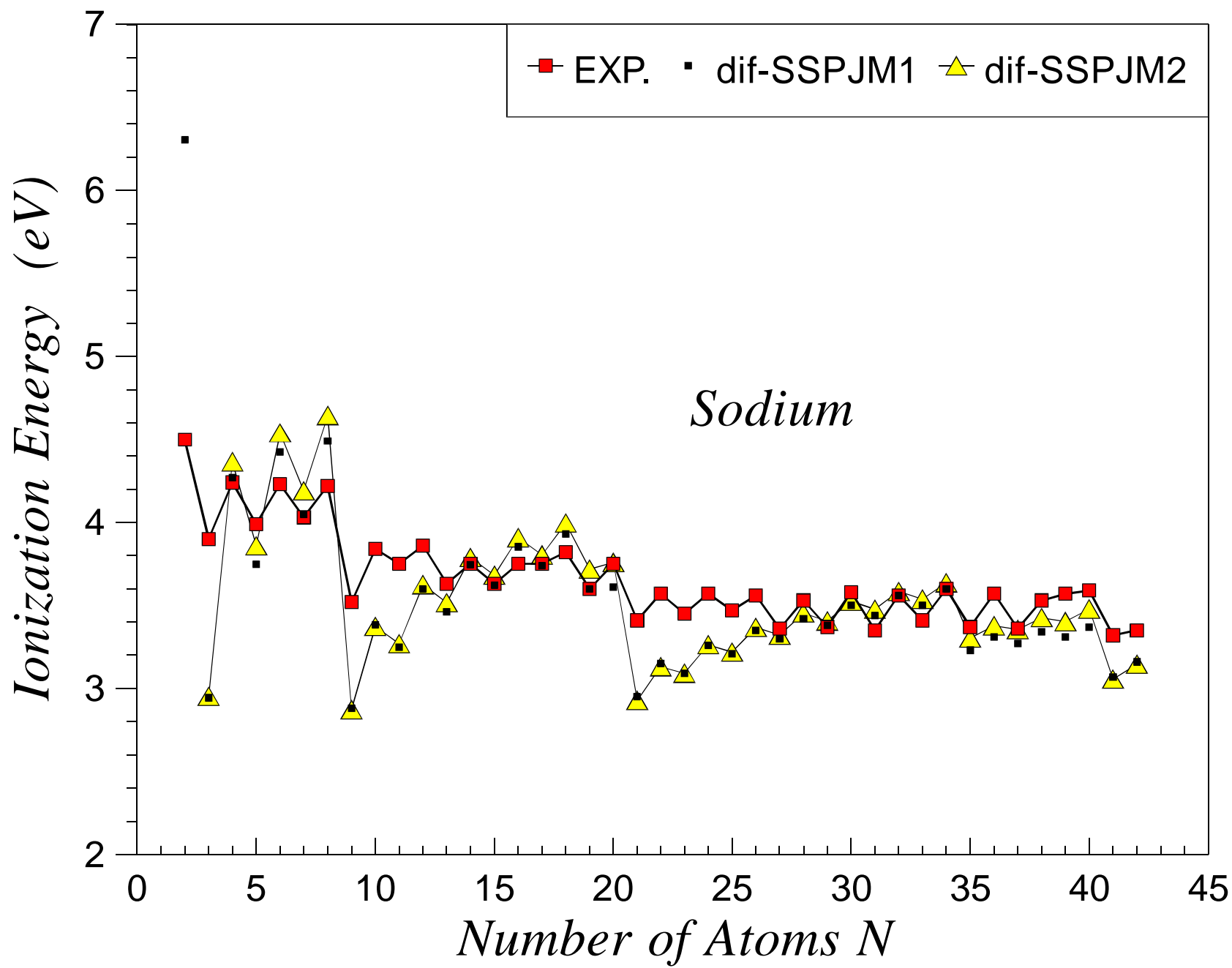
\end{document}